\documentclass{PoS}

\title{Broad absorption line (BAL) quasars as a class of low luminosity AGNs}

\ShortTitle{BALQSOs as a class of low luminosity AGNs}

\author{\speaker{M. Kunert-Bajraszewska}$^{1}$, M. Ceg\l{}owski$^{1}$ , C. Roskowi\'nski$^{1}$, M. Gawro{\'n}ski$^{1}$\\
        $^{1}$Toru\'n Centre for Astronomy, Faculty of Physics, Astronomy
and Informatics, NCU, Grudziacka 5, 87-100 Toru\'n, Poland\\
        E-mail: \email{magda@astro.uni.torun.pl}}

\abstract{Broad absorption lines seen in some quasars prove the existence of ionized plasma outflows from the accretion disk. Outflows together with powerful jets are important feedback processes. Understanding physics behind BAL outflows might be a key to comprehend Galaxy Evolution as a whole. First radio-loud BAL quasar was discovered in
1997 and this discovery has opened new possibilities for studies of the BAL phenomena, this time on the basis of radio emission. However, 
information about the radio structures, orientation and age of BAL quasars is still very limited due to weak radio emission and small sizes of these objects. 
Our high-resolution radio survey of a sample of BAL quasars aims to increase our knowledge about these objects. In this article, we present some conclusions arising from our research. }

\FullConference{12th European VLBI Network Symposium and Users Meeting\\
		7-10 October 2014\\
		Cagliari, Italy}

\begin{document}

\section{Introduction}
Jets and outflows are two main powerful processes transporting material out of the active galactic nuclei (AGN). Both are important feedback processes which means that they can efficiently interact with the surrounding medium or even self-regulate the growth of the supermassive black hole. Outflows are moving outward at lower speeds than jets but they can carry thousands of times more mass flux per unit of kinetic luminosity than collimated relativistic jets observed in $\sim10\%$ of all AGNs \cite{benoit, morabito}. We can observe them indirectly as blueshifted  Broad Absorption Lines (BALs) in quasar spectrum. However, only $\sim15\%$ of the whole population of quasars shows BALs in their spectra \cite{knigge2008}. This could be explained by the quasar unification scheme in which outflows are present in every quasar but appear as BALs in the spectrum only if they are seen under specific inclination \cite{elvis00}. 
Another explanation, which emerged with the discovery of radio-loud BAL quasars, suggests that absorption lines are only present in the early evolution phase of quasars \cite{becker00, gregg06}. 

Discovery of the existence of radio-loud BAL quasars gave us another opportunity to study
the BAL phenomenon, this time on the basis of radio emission. The radio emission is an
additional tool to understand their orientation and age by the VLBI imaging (detection of
radio jets and their direction, size determination), the radio-loudness parameter distribution
and variability study. The studies of the radio properties carried out by us on the sample of compact radio-loud BAL quasars brought some clues on the BAL phenomenon. We shortly summarize here the main conclusions coming from our work. More detailed discussion will be presented in forthcoming papers \cite{ceglowski}.

Throughout the paper, we assume a cosmology with
${\rm H_0}$=70${\rm\,km\,s^{-1}\,Mpc^{-1}}$, $\Omega_{M}$=0.3, $\Omega_{\Lambda}$=0.7. The adopted convention for the spectral index
definition is $S\propto\nu^{-\alpha}$.

\section{EVN and VLBI observations}
We performed high resolution radio observations of 26 compact radio-loud BAL quasars with EVN at 1.7\,GHz and with VLBA at 5 and 8.4\,GHz. The sources were selected by cross-matching {\it Faint Images of the Radio Sky at Twenty-cm (FIRST)} coordinates with he optical 
positions of BAL quasars from {\it A catalogue of Broad Absorption Line Quasars} from the {\it Sloan
Digital Sky Survey Third Data Release} made by \cite{trump06}. We limited our sample to objects with {\it FIRST} flux density more than 2 mJy and with side lobe probability less than 0.1. About $12\%$ of the identified BAL quasars appeared to have more than one radio counterparts within 60 arcseconds of SDSS position. We classified them as large-scale sources and excluded from the sample to avoid ambiguity in identification of the radio core that is important in our statistical studies. Finally our sample consisted of 309 compact radio-loud BAL quasars and 26 of them were chosen for further VLBI studies.

\begin{figure*}
\centering

   \includegraphics[width=0.31\textwidth, height=0.22\textheight]{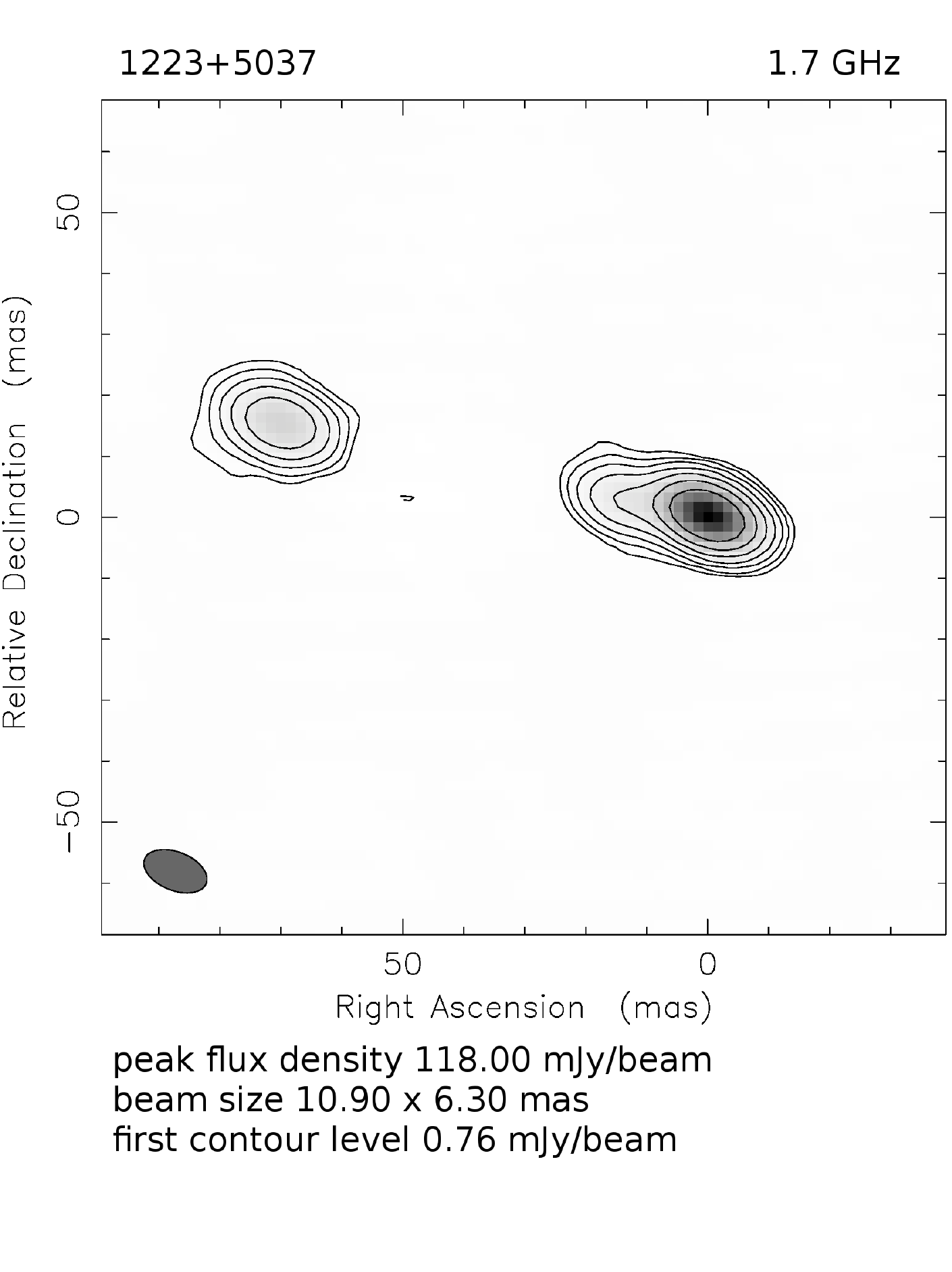}
   \includegraphics[width=0.31\textwidth, height=0.22\textheight]{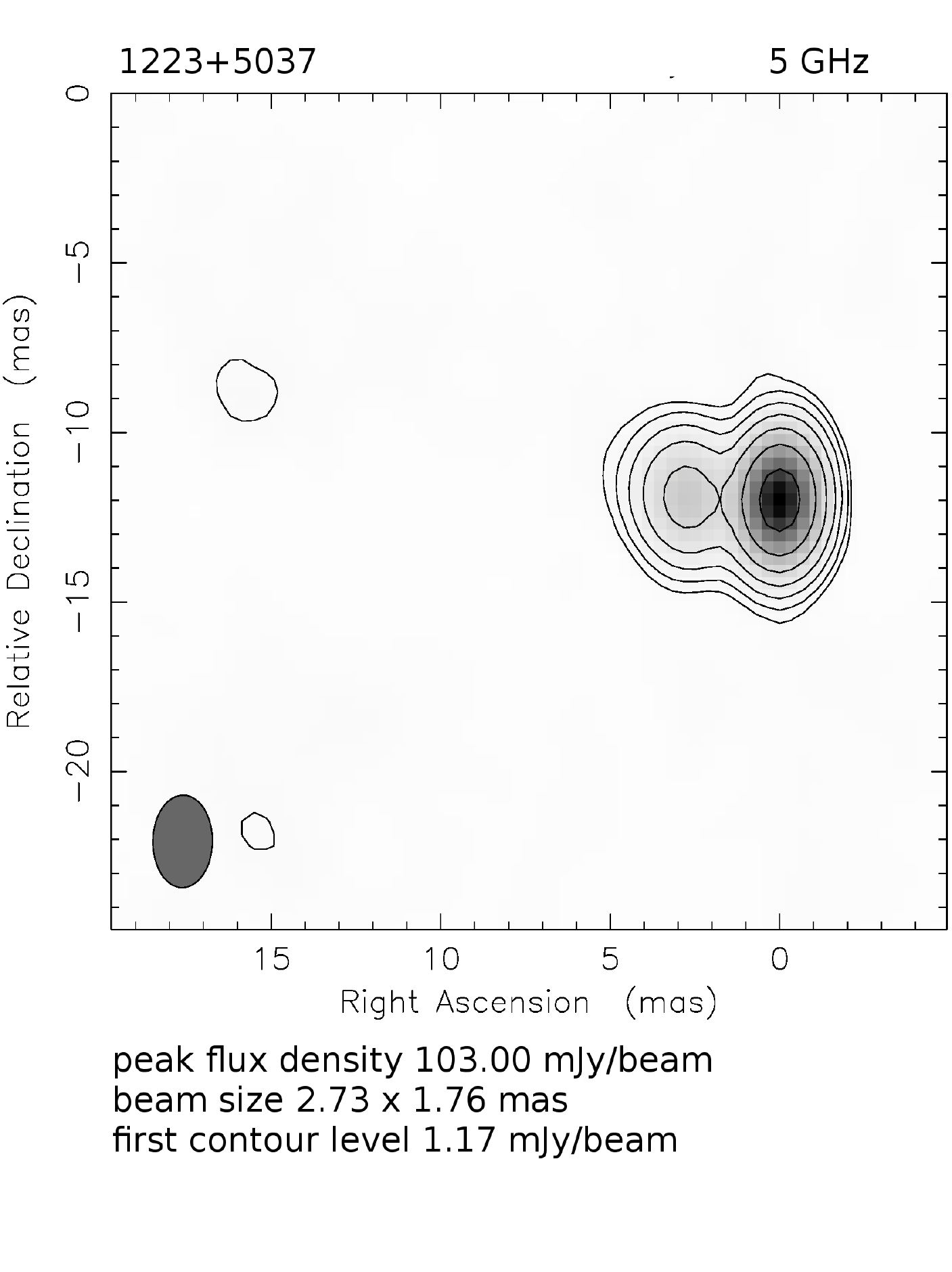}
   \includegraphics[width=0.31\textwidth, height=0.22\textheight]{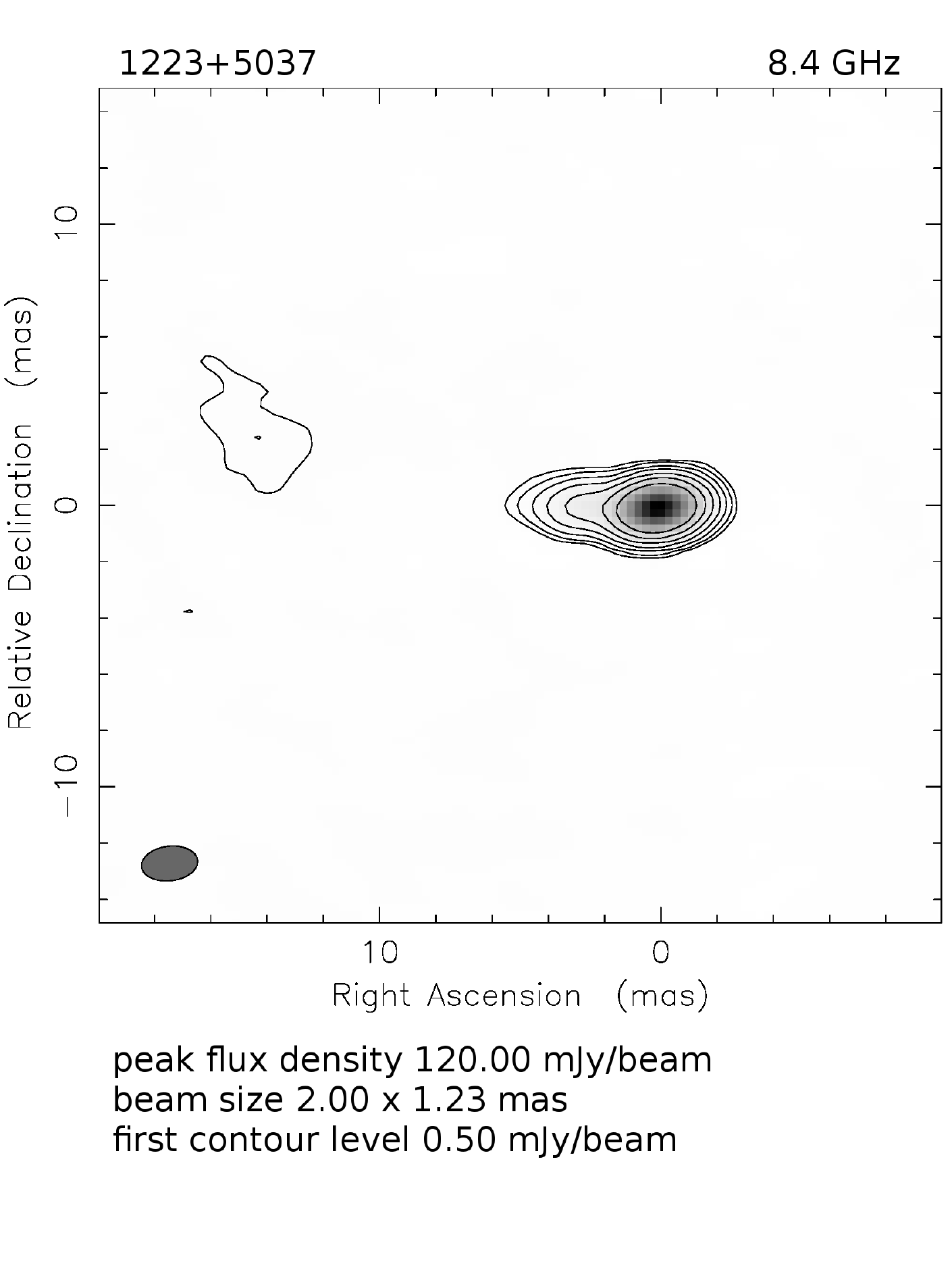}\\
   \includegraphics[width=0.31\textwidth, height=0.22\textheight]{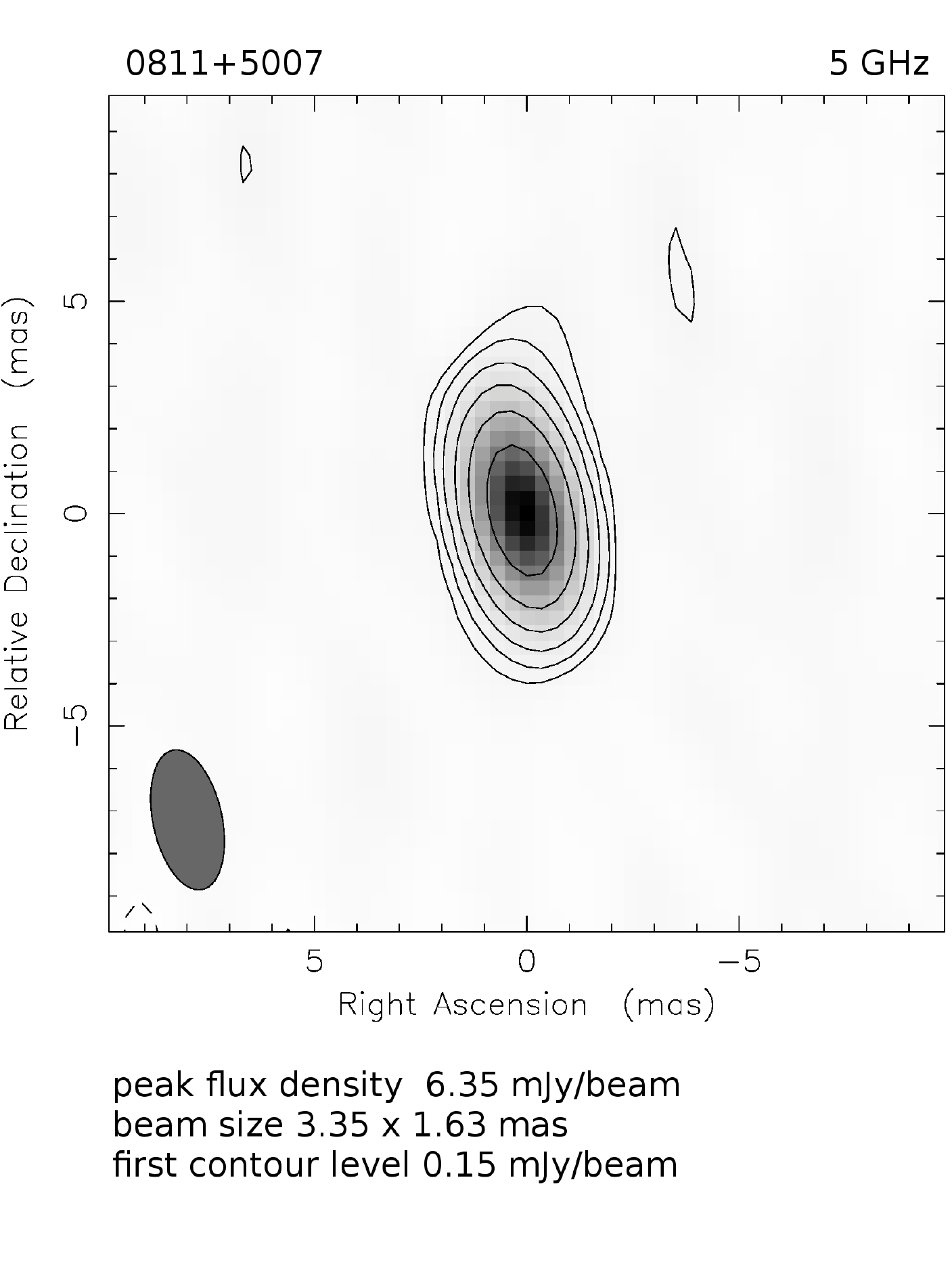}
   \includegraphics[width=0.31\textwidth, height=0.22\textheight]{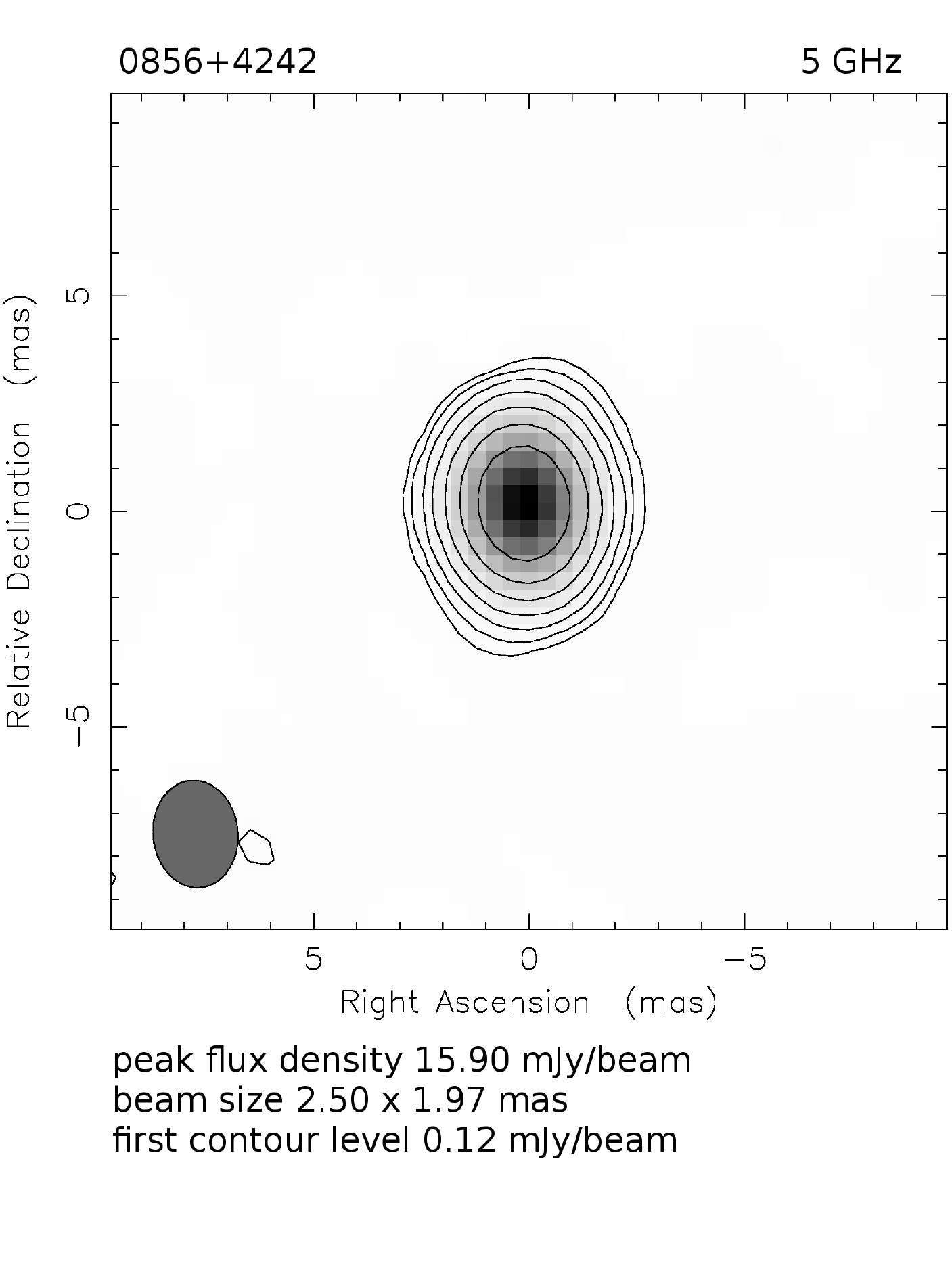}
   \includegraphics[width=0.31\textwidth, height=0.22\textheight]{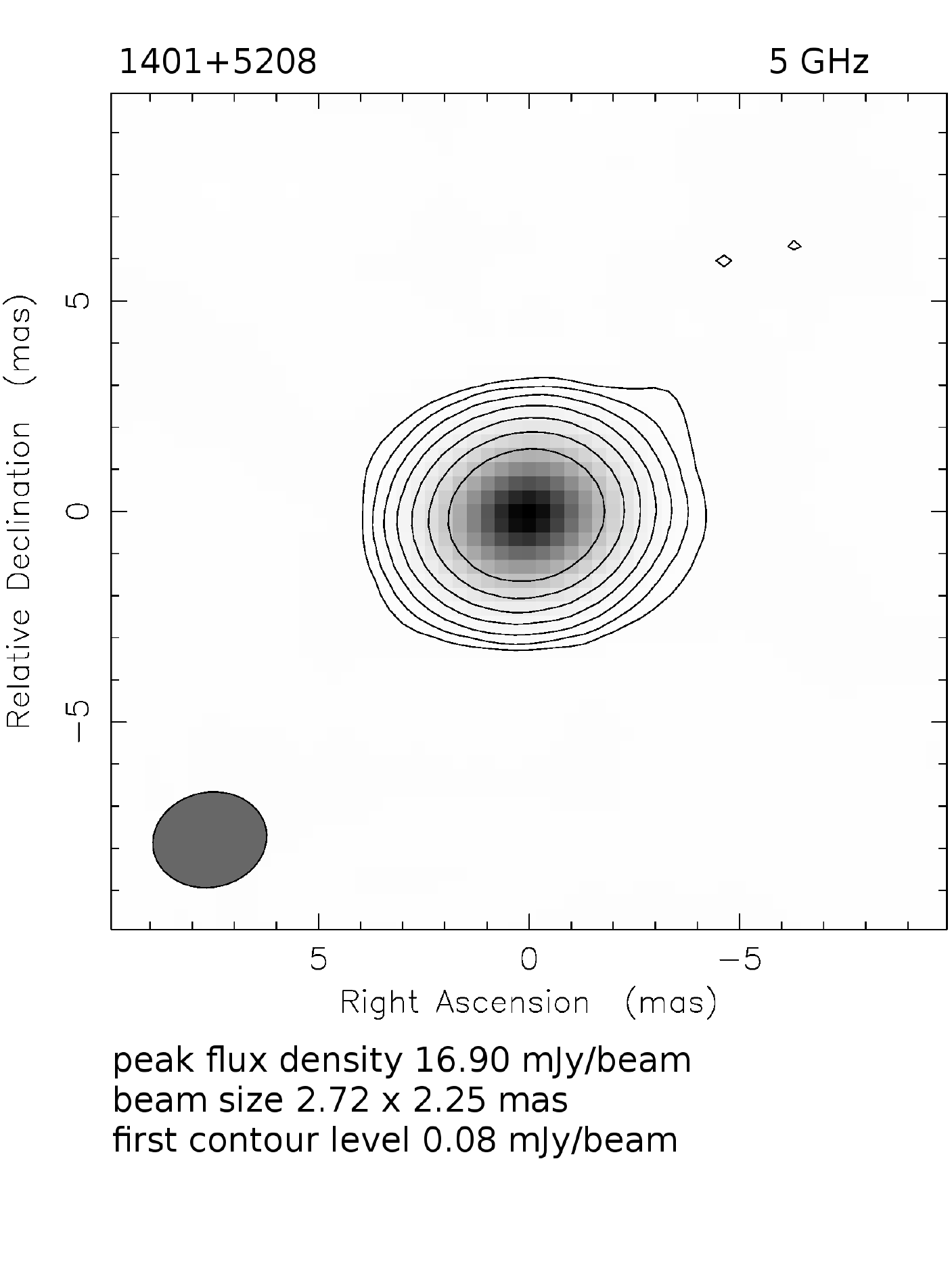}

\caption{Sources observed with EVN at 1.7\,GHz and VLBA at 5 and 8.5\,GHz. 
Contours increase by a factor 2, and the first contour level corresponds to $\approx 3\sigma$. Upper panel shows AI quasar 1223+5037 at three frequencies. Bottom panel shows three BI quasars observed with VLBA at 5\,GHz.}
\label{images}
\end{figure*}

It has to be noted here that the BAL definition is diffuse nowadays.
The first, now call traditional, way to quantify the BALs is via the balnicity index (BI) which account for
$\rm C_{IV}$ absorption troughs at least $\rm 2000\,km\,s^{-1}$ wide \cite{weymann91}.
However, this could potentially exclude the so-called mini-BALs with BI=0. Therefore, the more liberal absorption index (AI), including quasars
with weaker and much narrower absorption features (within $\rm 3000\,km\,s^{-1}$), was introduced by \cite{trump06}. Although the authors classify the SDSS sources as BAL quasars based on the absorption index, they also calculate the BI value of all their quasars. Therefore all sources from the catalogue in a 
natural way falls into two groups: 1) objects with AI$>$0 and BI=0, the mini-BALs (hereafter AI quasars) and 2) 
objects with AI$>$0 and BI$>$0, the so called 'true BAL quasars' (hereafter BI quasars). The optical and radio analysis of BAL quasars from both groups (AI$>$0 \& BI=0 and BI$>$0)
revealed many differences between them indicating that they constitute two independent classes \cite{shankar2008, knigge2008} and thus we treat them also as separate groups. Our final sample of 309 BAL quasars consists of 105 BI quasars and 204 AI objects.

For further high resolution VLBI observations we have selected comparably numbered samples of AI and BI sources with the largest 1.4\,GHz flux densities, a total of 26 objects which we called VLBI sample.
The AI quasars were observed in 2008 with EVN at 1.7\,GHz and with VLBA at 5 and 8.4\,GHz. The BI quasars were observed in 2013 with EVN and VLBA at 5\,GHz. Each target source was observed for approximately two hours on each frequency in phase-referencing mode in order to acquire radio maps with comparable dynamical range.  
In the case of EVN observations antennas in Jodrell Bank, Westerbork, Effelsberg, Onsala, Medicina, Toru\'n, Shanghai, Urumqi, Noto and Robledo took part in our experiment.

Data reduction including calibration and fringe-fitting was performed using Astronomical Image Processing System - AIPS package\footnote{http://www.aips.nrao.edu}. Imaging and self-calibration part
was performed using Difmap software \cite{shep93}. Some of the final images of the radio-loud BAL quasars are presented in Fig.~\ref{images}.

\section{Results}

During five observational campaigns (between 2008 and 2013),  92\% of the sources from VLBI sample were detected and successfully imaged. The multi-frequency observations of AI quasars provided us with spectral indices
and allowed for the correct classification of the sources. In the case of BI objects, observed only at 5\,GHz, we have made an estimations of their structures. We classified our sources in 3 categories: Single, Core-jet and Other. The unresolved sources were described as 'Single'. 'Core-jet' means the source has a bright central component and a one-sided jet. Last category 'Other' consists of objects with more complex morphologies. Majority of observed AI and BI quasars fall under 'core-jet' classification thus confirming the existence of non-thermal jets in BAL quasars. The upper panel of Fig.~\ref{images} shows the core-jet AI quasar 1223+5037 observed at three different frequencies. The bottom panel of Fig.~\ref{images} shows BI quasars 0811+5007, 0856+4242 and 1401+5208 classified as core-jet, single and core-jet, respectively (for more details see the proper regular publications). In the case of few objects we have noticed 
significant lack of flux density comparing to VLA or single dish observation. This is probably connected with more diffuse structures present in these sources which we did not fully detect in our observations due to their weakness. Previous publications show that the missing flux can also be explained by the re-activation scenario as the low-frequency remnant of the previous phase of the radio source activity \cite{bruni13,kun10a,kun10b,kun10}. There is a wide variety of spectral indices among BAL quasars. However, last publications show a hint that more often they possess steeper radio spectral index than non-BAL quasars \cite{dipompeo11, bruni13}. Most of the AI and BI quasars from our sample are steep spectrum objects. Only $\sim 12\%$ of the 26 BAL quasars observed by us have flat spectrum. Based on our radio observations we conclude that radio-loud BAL quasars represent
typical quasar geometries.

\begin{figure*}
\centering
   \includegraphics[width=0.47\textwidth, height=0.22\textheight]{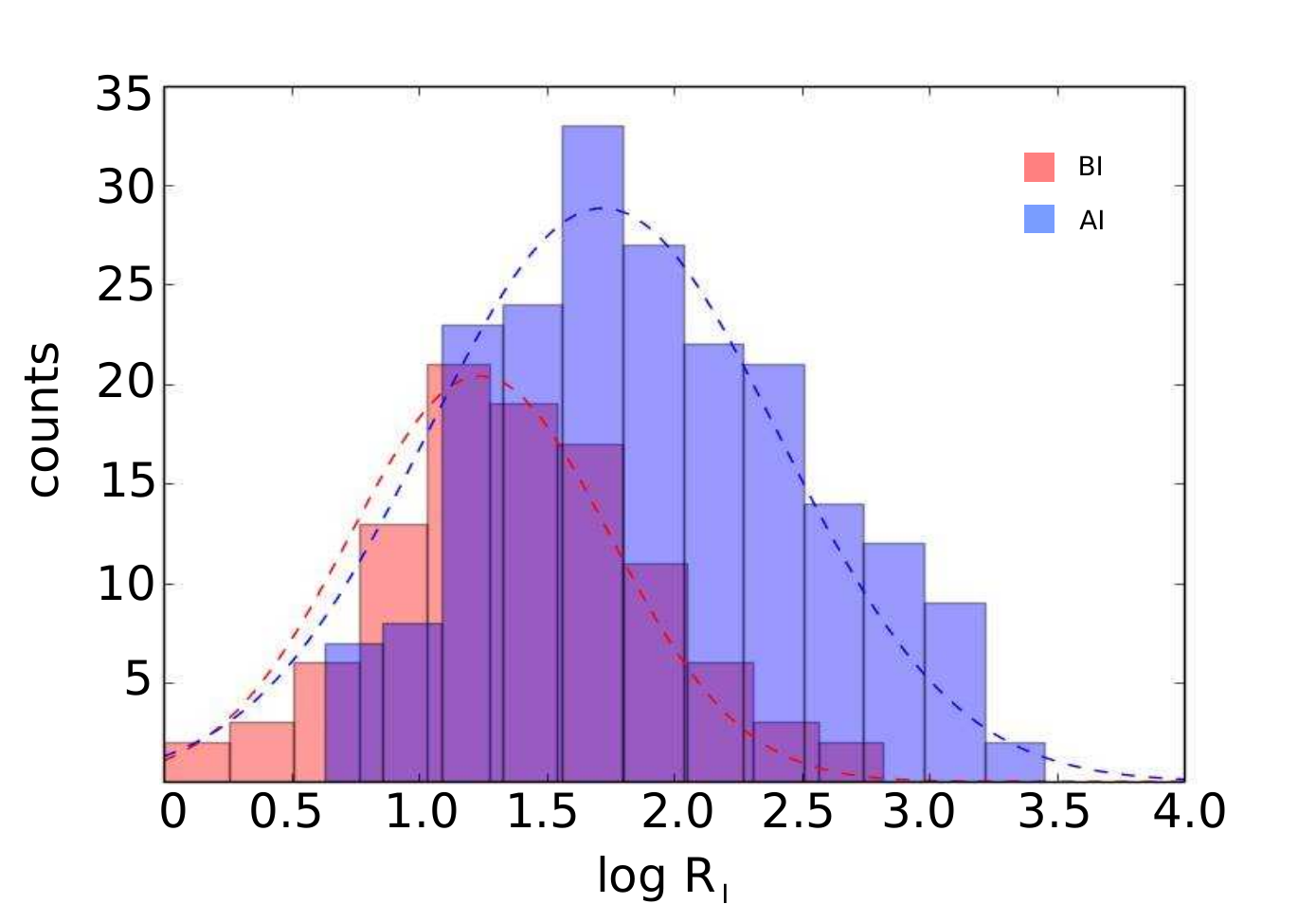}
   \includegraphics[width=0.52\textwidth, height=0.23\textheight]{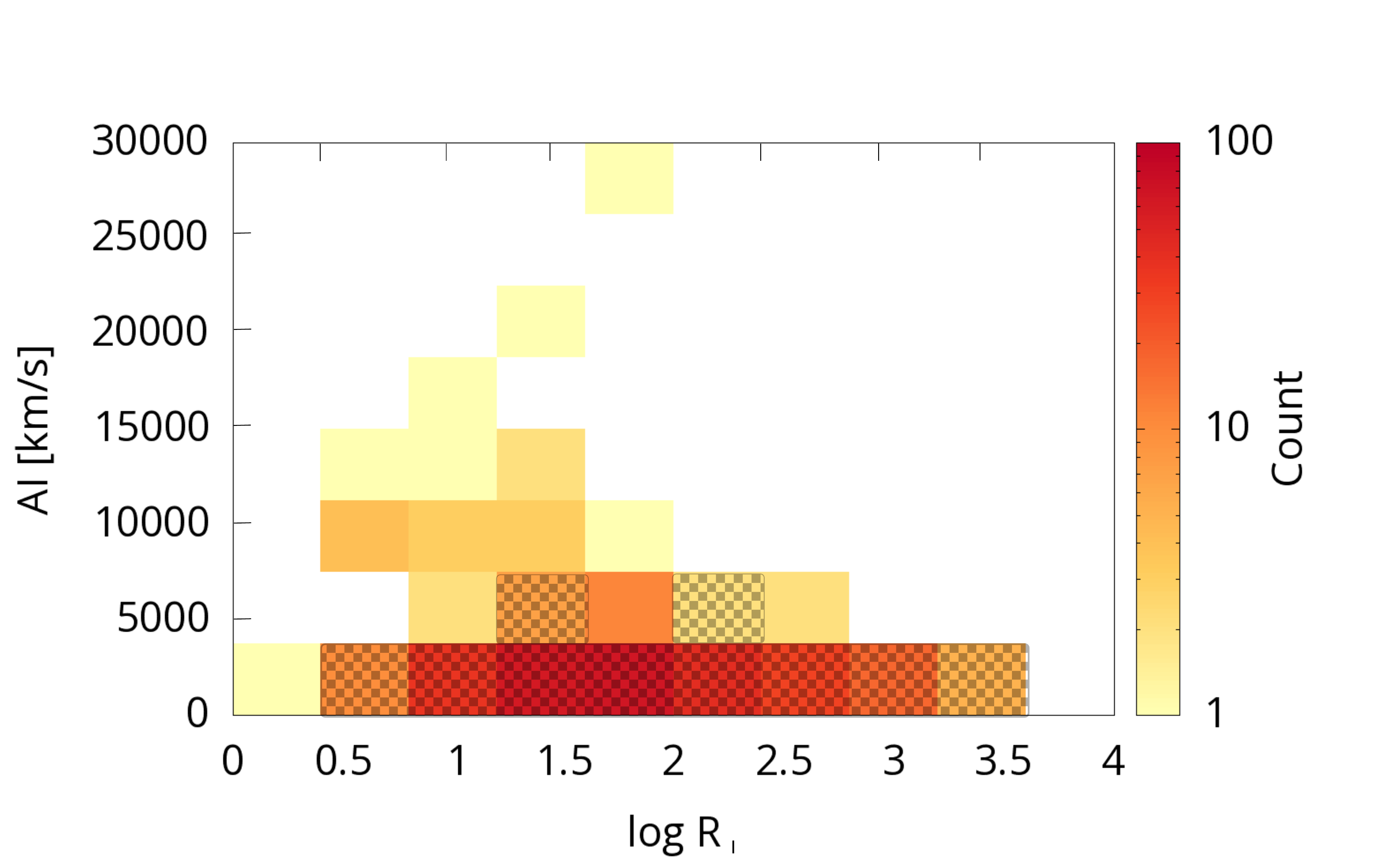}
\caption{Left panel: Radio-loudness, log\,$\rm R_{I}$, distribution for AI and BI quasars from parent sample. 
Gaussian function fitted to both distributions peaks at $\rm log\,R_{I}$ = 1.24 and 1.72 for BI and AI quasars, respectively. Right panel: Radio-loudness plotted versus the value of absorption index (AI) for the whole parent sample (309 sources). Hatched boxes indicate range of the AI values for AI quasars only.}
\label{counts}
\end{figure*}

Having a well selected large sample of compact radio-loud BAL quasars (309 objects) we have performed also statistical studies concerning orientation of these objects. Since the radio-to-optical ratio of the quasar
core, known as radio-loudness parameter, is thought to be a strong statistical indicator of orientation \cite{wills95} we examined its distribution among BAL quasars.  
We adopted radio-loudness definition from \cite{kimball2011}: $\rm R_{I} = (M_{radio}-M_{i})/-2.5$,
where $\rm M_{radio}$ is a K-corrected radio absolute magnitude and $\rm M_{i}$ is a Galactic reddening corrected and K-corrected i-band absolute magnitude. Since all our sources are unresolved on FIRST resolution we used their integrated 1.4\,GHz flux densities as a core flux and assume the radio core 
spectral index and optical spectral index to be 0 and -0.5 respectively. As discussed in previous section, the AI and BI quasars probably constitute different classes of objects. The result of Kolmogorov-Smirnov (K-S) test performed on the distribution of radio-loudness parameter $\rm R_{I}$ for BI and AI samples implies that both classes differ at the 0.05 confidence level. Histogram clearly hints that while $\rm R_{I}$ rises AI sample outnumbers BI (Fig.~\ref{counts}). As already mentioned, the $\rm R_{I}$ parameter is considered as an indicator of orientation and if this is the case its large numbers,
possibly $\rm log\,R_{I}>2.5$, could mean close to the radio jet axis orientation. Quasars
with $\rm log\,R_{I}<1.5$ are thought to be viewed at small angles relative to the plane of the disk \cite{kimball2011}. Thus, the radio-loudness distribution of BAL quasars may indicate that on average BI quasars are viewed closer to the disk plane than AI sources. Majority of BAL quasars observed by us in the VLBI technique (the 26 selected sources) have $\rm log\,R_{I}$ greater than 2 and 2.5 for BI and AI objects, respectively. This means that they belong to the tail of log\,$\rm R_{I}$ distribution
for both classes and constitute the most luminous subgroups of BAL quasars population.

Finally, we contrasted the radio-loudness values with the absorption index (AI) for the whole parent sample of BAL quasars (Fig.~\ref{counts}, right panel). It can be noticed that strong absorption, which is exclusively visible in BI quasars, is associated with lower values of radio-loudness parameter, log\,$\rm R_{I}<1.5$
and hence probably visible close to the disk plane. This relationship however, is not so obvious for radio stronger AI quasars, that may constitute independent group of objects. 

\section{Summary}

We performed multi-frequency high resolution radio observations of a sample of BAL quasars selected from the Sloan Digital Sky Survey (SDSS) DR3. Most of the sources were resolved showing, typical for radio-loud quasars, core-jet morphology. Their high radio luminosities and small linear sizes ($<$6\,kpc) indicate they are strong young AGNs. We then used the largest available sample of BAL quasars to study the relationships between the radio and optical properties in these objects. We found that the stronger absorption is connected with the lower values of the radio-loudness parameter, log\,$\rm R_{I}$, and thus probably with larger viewing angles. Therefore, the orientation is an important parameter that affects the amount of the measured absorption. Nevertheless the lack of correlation between the value of absorption and parameter $\rm R_{I}$ suggest that some additional factors may be significant here, e.g. weak radio emission. 

\section{Acknowledgments}

This work was supported by the National Scientific Centre under grant DEC-2011/01/D/ST9/00378.
The research leading to these results has received funding from the European Commission Seventh Framework Programme (FP/2007-2013) under grant agreement No 283393 (RadioNet3).

\end{document}